\documentclass[a4paper,onecolumn,superscriptaddress,11pt,accepted=2018-07-30]{quantumarticle}
\pdfoutput=1
\usepackage[utf8]{inputenc}
\usepackage[english]{babel}
\usepackage[T1]{fontenc}
\usepackage{amsmath}
\usepackage{hyperref}

\usepackage{tikz}
\usepackage{lipsum}

\begin{document}

\title{Quantum Computing in the NISQ era and beyond}
%\date{\today}
\date{30 July 2018}
\author{John Preskill}
\affiliation{Institute for Quantum Information and Matter and Walter Burke Institute for Theoretical Physics, \\ 
California Institute of Technology, Pasadena CA 91125, USA}
\email{preskill@caltech.edu}
\thanks{Based on a Keynote Address at Quantum Computing for Business, 5 December 2017.}
%\homepage{http://quantum-journal.org}
%\orcid{0000-0003-0290-4698}
%\thanks{You can use the \texttt{\textbackslash{}email}, \texttt{\textbackslash{}homepage}, and \texttt{\textbackslash{}thanks} commands to add additional information for the preceding \texttt{\textbackslash{}author}. If applicable, this can also be used to indicate that a work has previously been published in conference proceedings.}

\maketitle

\begin{abstract}
Noisy Intermediate-Scale Quantum (NISQ) technology will be available in the near future. Quantum computers with 50-100 qubits may be able to perform tasks which surpass the capabilities of today's classical digital computers, but noise in quantum gates will limit the size of quantum circuits that can be executed reliably. NISQ devices will be useful tools for exploring many-body quantum physics, and may have other useful applications, but the 100-qubit quantum computer will not change the world right away --- we should regard it as a significant step toward the more powerful quantum technologies of the future. Quantum technologists should continue to strive for more accurate quantum gates and, eventually, fully fault-tolerant quantum computing.
%%%
%\begin{center}
%\textit{Based on a Keynote Address at Quantum Computing for Business, 5 December 2017~\footnote{See  {\tt q2b.us} for %videos of all the talks at the meeting.}}
%\end{center}
\end{abstract}

\section{Introduction}

%The premise of this meeting is that now is an opportune time for a fruitful discussion among researchers, entrepreneurs, managers, and investors who share an interest in quantum computing. 
%%%
Now is an opportune time for a fruitful discussion among researchers, entrepreneurs, managers, and investors who share an interest in quantum computing. There has been a recent surge of investment by both large public companies and startup companies, a trend that has surprised many quantumists working in academia. While we have long recognized the commercial potential of quantum technology, this ramping up of industrial activity has happened sooner and more suddenly than most of us expected. 

%I'm pleased by this chance to participate in assessing the current status and future potential of quantum computing. 
In this article I assess the current status and future potential of quantum computing. 
Because quantum computing technology is so different from the information technology we use now, we have only a very limited ability to glimpse its future applications, or to project when these applications will come to fruition. While this uncertainty fuels optimism, our optimism should be tempered with caution. We may feel confident that quantum technology will have a substantial impact on society in the decades ahead, but we cannot be nearly so confident about the commercial potential of quantum technology in the near term, say the next five to ten years. That is the main message I hope to convey. 
%% in this talk. 
%%%
With that said, I'm sure that vigorous discussion among all the interested parties
%, like that facilitated by this meeting, 
can help light the way toward future progress. 

\section{Opportunities at the entanglement frontier}
%%
%I'm excited to be here and happy to be part of fostering the collaborative spirit of our community. 

I am a theoretical physicist with a background in particle physics and cosmology, but for more than 20 years much of my research effort has been directed toward quantum information science. I'm drawn to this field because I feel we are now in the early stages of exploring a new frontier of the physical sciences, what we might call the \textit{complexity frontier} or the  \textit{entanglement frontier}. This new frontier, while different than the frontier of particle physics or cosmology, is very fundamental and exciting. Now, for the first time in human history, we are acquiring and perfecting the tools to build and precisely control very complex, highly entangled quantum states of many particles, states so complex that we can't simulate them with our best digital computers or characterize them well using existing theoretical tools. This emerging capability will open the door to new discoveries.

For a physicist like me, what is really exciting about quantum computing is that we have good reason to believe that a quantum computer would be able to efficiently simulate any process that occurs in Nature. We don't think that's not true for classical ({\textit{i.e.}, non-quantum) digital computers, which (as far as we know) can't simulate highly entangled quantum systems. With quantum computers we should be able to probe more deeply into the properties of complex molecules and exotic materials, and also to explore fundamental physics in new ways, for example by simulating the properties of elementary particles, or the quantum behavior of a black hole, or the evolution of the universe right after the big bang.

Our confidence that exploration of the entanglement frontier will be rewarding rests largely on two principles: (1) quantum complexity (our basis for thinking that quantum computing is powerful), and (2) quantum error correction (our basis for thinking that quantum computers are scalable to large devices solving hard problems). Underlying both of these principles is the idea of \textit{quantum entanglement}. Entanglement is the word we use for the characteristic correlations among the parts of a quantum system, which are quite different from the correlations we encounter in everyday life. To understand the concept of entanglement, imagine a system with many parts, for example a book which is 100 pages long. For an ordinary classical 100-page book, every time you read another page you learn another 1\% of the content of the book, and after you have read all of the pages one by one you know everything that's in the book. But now suppose instead that it's a quantum book, where the pages are very highly entangled with one another. Then when you look at the pages one at a time you see only random gibberish, and after you have read all the pages one by one you know very little about the content of the book. That's because the information in the quantum book is not imprinted on the individual pages; it is encoded almost entirely in how the pages are correlated with one another. If you want to read the book you have to make a collective observation on many pages at once. That's quantum entanglement, the essential feature making information carried by quantum systems very different from information processed by ordinary digital computers. 

\section{The potential of quantum computing}

\subsection{Why we think quantum computing is powerful}
An ordinary computer processes bits, where each bit could be, say, a switch which is either one or off. But to build highly complex entangled quantum systems, the fundamental information-carrying components of a quantum computer must be quantum bits, what we call \textit{qubits}. A qubit can be realized physically in many different ways. It can be carried by a single atom, or a single electron, or a single photon (a particle of light). Or a qubit can be carried by a more complicated system, like a very cold superconducting electrical circuit in which many electrons are moving. 

%For a typical very highly entangled system of just a few hundred qubits, describing completely all the correlations among those qubits would require writing down a number of bits which is larger than the number of atoms in the visible universe. It will never be possible to write such a description down, and that opens the possibility that by processing qubits we can perform tasks that wouldn't be possible by processing bits. 

When we speak of quantum complexity, what springs to mind is the staggering complexity of using ordinary classical data to describe highly entangled quantum states of many qubits. Giving a complete description of all the correlations among just a few hundred qubits may require more bits than the number of atoms in the visible universe. It will never be possible, even in principle, to write that description down, or to faithfully describe the processing of a few hundred qubits using classical language. 

This apparently extravagant complexity of the quantum world, though highly suggestive, does not by itself ensure that quantum computers are more powerful than classical ones. 
%For one thing, to read out the result of a quantum computation we need to measure the qubits, and a single-shot measurement of $n$ qubits can collect no more than $n$ bits of information. This means that in order to access all the correlations among the qubits we would need to prepare and measure the $n$-qubit state a vast number of times. Furthermore, the quantum states that can be prepared in a quantum computer which runs for a reasonable amount of time  perhaps a less extravagant description would suffice if the preparation and preparation is repeated a more modest number of times. 
%%
But we have at least three good reasons for thinking that quantum computers have capabilities surpassing what classical computers can do. 

\begin{description} 
\item (1) \textit{Quantum algorithms for classically intractable problems.}
First, we know of problems that are believed to be hard for classical computers, but for which quantum algorithms have been discovered that could solve these problems easily. The best known example is the problem of finding the prime factors of a large composite integer \cite{shor}. We believe factoring is hard because many smart people have tried for many decades to find better factoring algorithms and haven't succeeded. Perhaps a fast classical factoring algorithm will be discovered in the future, but that would be a big surprise. 

\item (2) \textit{Complexity theory arguments.} The theoretical computer scientists have provided arguments, based on complexity theory, showing (under reasonable assumptions) that quantum states which are easy to prepare with a quantum computer have superclassical properties; specifically, if we measure all the qubits in such a state we are sampling from a correlated probability distribution that can't be sampled from by any efficient classical means \cite{bremner,harrow-montenaro}. %indicating that the quantum computer is doing things that surpass what we can do classically. 

\item (3) \textit{No known classical algorithm can simulate a quantum computer.} But perhaps the most persuasive argument we have that quantum computing is powerful is simply that we don't know how to simulate a quantum computer using a digital computer; that remains true even after many decades of effort by physicists to find better ways to simulate quantum systems. 

\end{description}

It's a remarkable claim --- one of the most amazing ideas I've encountered in my scientific life --- that there is a distinction between problems that are classically hard and problems that are quantumly hard. And it is a compelling challenge to understand better what problems are classically hard but quantumly easy \cite{zoo,montenaro}. We should recognize in particular that the power of a quantum computer is not unlimited. We don't expect, for example, that a quantum computer will be able to solve efficiently the hard instances of NP-hard problems like the traveling salesman problem. For such hard combinatorial search problems we probably can't do much better than exhaustively searching for a solution. Quantum computers can speed up exhaustive search \cite{grover}, but only modestly \cite{bbbv}, so NP-hard problems are likely to be quantumly hard as well as classically hard.

For a physicist seeking problems which are classically hard and quantumly easy, the natural place to look is the task of simulating a many-particle quantum system. As two great physicists, Bob Laughlin and David Pines, put it some years ago \cite{laughlin}, we have a ``theory of everything that is relevant to ordinary life.'' We have high confidence this theory is correct, and we can write down the equations precisely ---  they are the equations that describe how atomic nuclei and electrons interact electromagnetically. But we can't solve those equations. And so as Laughlin and Pines put it: ``We have a theory of everything only to discover that it has revealed exactly nothing about many things of great importance.'' Those things of importance they envisaged are the situations in the quantum world where entanglement has profound consequences. Dramatizing the futility of the task they proclaimed: ``No computer existing, or that will ever exist, can break this barrier'' of solving the equations describing many entangled particles.

But in fact, years before Laughlin and Pines wrote these words, the physicist Richard Feynman had articulated a rebuttal \cite{feynman}. As Feynman put it: ``Nature isn't classical dammit, and if you want to make a simulation of Nature you better make it quantum mechanical, and by golly it's a wonderful problem because it doesn't look so easy.'' Feynman had envisioned using a quantum computer to solve the quantum physics problems that physicists and chemists had failed to solve using digital computers. Laughlin and Pines knew well that Feynman had made this proposal years earlier, but had dismissed his idea as impractical. Now, some 35 years after Feynman's proposal, we're just beginning to reach the stage where quantum computers can provide useful  solutions to hard quantum problems. 

\subsection{Why quantum computing is hard}
So why is it taking so long? What is it about quantum computing that's so difficult? The core of the problem stems from a fundamental feature of the quantum world --- that we cannot observe a quantum system without producing an uncontrollable disturbance in the system. That means that if we want to use a quantum system to store and reliably process information, then we need to keep that system nearly perfectly isolated from the outside world. At the same time, though, we want the qubits to strongly interact with one another so we can process the information; we also need to be able to control the system from the outside, and eventually read out the qubits so we can find the result of our computation. It is \textit{very} challenging to build a quantum system that satisfies all of these desiderata. It has taken many years of development in materials and control and fabrication to get where we are now. 

Eventually we expect to be able to protect quantum systems and scale up quantum computers using the principle of quantum error correction \cite{gottesman}. The essential idea of quantum error correction is that if we want to protect a quantum system from damage then we should encode it in a very highly entangled state; like that 100-page book I described earlier, this entangled state has the property that the environment, interacting with parts of the system one at a time, is unable to glimpse the encoded information and therefore can't damage it. Furthermore, we've understood in principle how to process quantum information which is encoded in a  highly entangled state. Unfortunately, there is a significant overhead cost for doing quantum error correction --- writing the protected quantum information into a highly entangled book requires many additional physical qubits --- so reliable quantum computers using quantum error correction are not likely to be available very soon. 

\section{The NISQ era unfolds}

\subsection{The 50-qubit barrier}

%%%
%% [Summary of the state of the art of hardware.]
%%%

Even with fault-tolerant quantum computing still a rather distant dream, we are now entering a pivotal new era in quantum technology. For this talk, I needed a name to describe this impending new era, so I made up a word: {\textit{NISQ}. This stands for \textit{Noisy Intermediate-Scale Quantum}. Here ``intermediate scale'' refers to the size of quantum computers which will be available in the next few years, with a number of qubits ranging from 50 to a few hundred.\footnote{72-qubit and 50-qubit devices, based on superconducting circuits, have been announced recently by Google and IBM, respectively.} 50 qubits is a significant milestone, because that's beyond what can be simulated by brute force using the most powerful existing digital supercomputers.\footnote{This ``milestone'' is actually a bit fuzzy, in several respects. In particular, the resources needed in a classical simulation depend on the depth (number of time steps) of the quantum circuit, as well as on the number of qubits \cite{boixo,chen,ibm}.} ``Noisy'' emphasizes that we'll have imperfect control over those qubits; the noise will place serious limitations on what quantum devices can achieve in the near term. 

Physicists are excited about this NISQ technology, which gives us new tools for exploring the physics of many entangled particles. It might also have useful applications of interest to 
%this audience 
the business community, but we're not sure about that. We shouldn't expect NISQ is to change the world by itself; instead it should be regarded as a step toward more powerful quantum technologies we'll develop in the future. I do think that quantum computers will have transformative effects on society eventually, but these may still be decades away. We're just not sure how long it's going to take. 

\subsection{Qubit ``quality''}

I've emphasized the number of qubits as a measure of how difficult it is to do the simulation of a quantum computer on a classical device, but the number of qubits isn't the only thing we care about. We also care about the \textit{quality} of the qubits, and in particular the accuracy with which we can perform quantum gates --- well-controlled entangling operations acting on pairs of qubits. With the best hardware we have now for controlling trapped ions \cite{lucas} or superconducting circuits \cite{martinis}, the error rate per gate for two-qubit gates is above the $.1$\% level (and often much worse). Furthermore, we don't yet know whether error rates that low can be maintained in larger devices with many qubits; perhaps we'll find out soon. Naively, then, and as I'll say later this might be too naive, with these noisy devices we don't expect to be able to execute a circuit that contains many more than about 1000 gates --- that is, 1000 fundamental two-qubit operations --- because the noise will overwhelm the signal in a circuit much larger than that. 
That limitation on circuit size imposes a ceiling on the computational power of NISQ technology. Eventually we'll do better, using quantum error correction to scale up to larger circuits. But as I've already emphasized, since quantum error correction imposes a heavy overhead cost in number of qubits and number of gates, scaling up using quantum error correction is a more distant goal. When I speak of the NISQ era, I'm imagining quantum computers with noisy gates unprotected by quantum error correction. 

There are other things we care about, too, aside from the number of qubits and the gate error rate. The time it takes to execute a single  gate is also important for setting the time scale needed for a quantum computer to solve a problem, and it is noteworthy that superconducting circuits are about a thousand times faster than ion trap quantum processors. We need to be able to prepare and measure qubits accurately; currently the measurement error probability is about $1\%$ for superconducting qubits, and much better than that for trapped ions. We care about the connectivity among the qubits --- for which pairs of qubits in our device can we perform an accurate two-qubit gate? It is also important to know how reliably we can fabricate qubits --- if we attempt to construct a many-qubit device, how many of the qubits actually perform well enough to be useful? These considerations and others, too, should be kept in mind when we make comparisons across different quantum computing platforms.

\section{What I won't say much about}

Before moving on to the main focus of this article, I want to include a disclaimer to avoid misunderstanding. There are some very interesting and important aspects of quantum technology that I'm not really going to talk about, but I'll mention them so you'll know I'm not going to talk about them. 

\begin{description}
 \item (1) \textit{Quantum-resistant cryptography.} We foresee disruptive effects of quantum computers on how we protect our privacy --- the public key cryptosystems that are in widespread use today will become obsolete in the years ahead because they can be easily broken by sufficiently powerful quantum computers. We should be thinking now about how we're going to protect our privacy in the future. One possible way is to replace our existing cryptosystems with new ones which we're confident are resistant to attacks by quantum computers \cite{bernstein}.  

\item (2) \textit{Quantum key distribution, quantum networks, and quantum repeaters.} We might also use traveling qubits --- most likely photons --- to create shared keys that can be used for encryption and decryption. Quantum key distribution exploits the principle that you cannot eavesdrop on quantum communication without producing a detectable disturbance \cite{qkd}. But we don't yet have the technology to distribute quantum entanglement, and hence secret key, around the world; that's another interesting technological challenge \cite{long-distance}. It is also notable that, aside from applications to key exchange, a global quantum network might be used for other purposes, like sharing information among quantum devices. 

\item (3) \textit{Quantum randomness expansion.} Unlike deterministic classical devices, quantum devices can generate intrinsic randomness. Remarkably, this feature of quantum physics can be exploited to expand a short random seed to a much longer string of \textit{certifiably} random bits, even if one distrusts the equipment used for this purpose --- it suffices to assume that spacelike separated parties are unable to communicate with one another \cite{nist-randomness}, or to make suitable assumptions about (quantum) computational hardness \cite{vidick-random}. Certifiable randomness has many potential applications, for example to secure communication protocols, unbiased statistical sampling, and Monte Carlo simulations. 

\item (4) \textit{Quantum sensing.} Quantum technology has advantages for some kinds of sensing. Quantum systems can sense weak forces with better sensitivity and higher spatial resolution than other sensing technologies \cite{sensing}; thus quantum sensing could have relatively near-term high-impact applications, to medicine for example. 

\end{description}

Quantum communication, networking, randomness expansion, and sensing cannot be cleanly separated from quantum computing, because similar technological challenges are faced by all these quantum-information goals. Nevertheless, I'll give short shrift to %quantum cryptography, networking, and sensing 
these topics here. 

\section{Quantum speedups?}

What I would like to focus on instead is whether quantum computers will have widely used applications, particularly in the relatively near term. The main question is: When will quantum computers be able to solve problems we care about faster than classical computers, and for what problems?

At least in the near term quantum computers are likely to be special purpose devices, which most users will access via the cloud. When we speak of a quantum speedup, we typically mean that the quantum computer solves the problem faster than competing classical computers using the best available hardware and running the best algorithm which performs the same task. (Arguably, though, quantum technology might be preferred even if classical supercomputers run faster, if for example the quantum hardware has lower cost and lower power consumption.) In any case, we should recognize that the power of classical computers will continue to increase, with exascale systems (surpassing $10^{18}$ FLOPS) expected to be available in a few years. Quantum computers are striving to catch up with a moving target, as both classical hardware and classical algorithms continually improve. 

A few years ago I spoke enthusiastically about \textit{quantum supremacy} as an impending milestone for human civilization \cite{preskill}. I suggested this term as a way to characterize computational tasks performable by quantum devices, where one could argue persuasively that no existing (or easily foreseeable) classical device could perform the same task, disregarding whether the task is useful in any other respect. I was trying to emphasize that now is a very privileged time in the coarse-grained history of technology on our planet, and I don't regret doing so. But from a commercial perspective, obviously we should pay attention to whether the task is useful! Quantum supremacy is a worthy goal, notable for entrepreneurs and investors not so much because of its intrinsic importance but rather as a sign of progress toward more valuable applications further down the road. 

We should also bear in mind that, because of the imperfect performance of NISQ technology, it may be hard to validate that a quantum computer is really giving the right answer. That's particularly true for the quantum simulation problems physicists are excited about. So it's important for researchers to continue seeking better methods for verifying the output of a quantum computer.

%[Cautionary tales on classical algorithms.]

%[Quantum supremacy.] 

\subsection{Quantum optimizers}

I have already emphasized that we don't expect quantum computers to be able to solve efficiently worst-case instances of NP-hard problems like combinatorial optimization problems; still, it's conceivable (though not guaranteed) that quantum devices will be able to find better approximate solutions or find such approximate solutions faster. For example, we might formulate $m$ %easily computable 
constraints on $n$ bits, and seek an $n$-bit string which solves as many of the $m$ constraints as possible. We may say we have solved this optimization problem exactly if we can find the maximal number $k$ of constraints that can be satisfied simultaneously, and that we have an approximate solution if we can guarantee that the maximal number of satisfiable constraints is at least $k'$ for some $k' < k$;  the \textit{approximation ratio} $k'/k$ is a measure of the quality of our approximate solution.

For some optimization problems, even finding an approximation solution is NP-hard if the approximation ratio is sufficiently close to one \cite{khot}, and in those cases we don't expect a quantum computer to be able to find the approximate solution efficiently for hard instances of the problem. But for many problems there is a big gap between the approximation achieved by the best classical algorithm we currently know and the barrier of NP-hardness. So it would not be shocking to discover that quantum computers have an advantage over classical ones for the task of finding approximate solutions, an advantage that some users might find quite valuable. 

Whether that quantum advantage really exists is currently an open question, but we will soon have the opportunity to explore the question experimentally using quantum hardware. Of course, even if there is such a quantum advantage for approximate optimization, NISQ technology might be inadequate for demonstrating the advantage; still, it will be fun to try it and see how well it works. 

The emerging paradigm for solving optimization problems using near-term quantum technology is a kind of hybrid quantum-classical algorithm. In this scheme, we use a quantum processor to prepare an $n$-qubit state, then measure all the qubits and process the measurement outcomes using a classical optimizer; this classical optimizer instructs the quantum processor to alter slightly how the $n$-qubit state is prepared. This cycle is repeated many times until it converges to a quantum state from which the approximate solution can be extracted. When applied to classical combinatorial optimization problems, this procedure goes by the name Quantum Approximate Optimization Algorithm (QAOA) \cite{qaoa}. But it can also be applied to quantum problems, like finding low-energy states of many-particle quantum systems (large molecules, for example). When applied to quantum problems this hybrid quantum-classical procedure goes by the name Variational Quantum Eigensolver (VQE) \cite{vqe}.

Will NISQ technology running QAOA or VQE be able to outperform classical algorithms that find approximate solutions to the same problems? Nobody knows, but we're going to try it and see how well we can do. It's an ambitious goal, because the classical methods we use to solve these problems are well honed after decades of development. On the other hand, even if early-generation NISQ devices are not yet ready to compete with the best classical computers, experimental results might encourage our hopes that QAOA and/or  VQE can surpass classical methods in the future, and so spur further technological improvements.

\subsection{How quantum testbeds might help}
The history of classical computing teaches us that when hardware becomes available that stimulates and accelerates the development of new algorithms. There are many examples of heuristics that were discovered experimentally, which worked better than theorists could initially explain. We can anticipate that the same thing will happen with quantum computers.

For example, theorists eventually explained why the simplex method for linear programming works well in practice \cite{spielman}, but only long after it had been found to be useful experimentally. A current example is deep learning \cite{hinton}; we lack a good theoretical explanation for why it works as well as it does.  In the quantum case, too, experiments may validate the performance of heuristic algorithms, where we don't understand why they work. That might happen in particular for the quantum optimization algorithms we're planning to test. 

But I emphasize again that the imperfect gates in NISQ systems will severely limit their computational power, because circuits with many gates will be too noisy to give useful results.  Near-term experiments will explore what we can do with of order 100 qubits, and with circuit depth (number of time steps) less than 100, maybe much less. A vibrant discussion between quantum algorithm designers and application users might point those experiments in promising directions; %an important goal for a meeting like this one is 
I hope that this article can help to facilitate and inspire that discussion.

\subsection{Quantum annealing}

I have emphasized 50 to 100 qubits as a coming milestone for quantum technology, but in fact we have a 2000-qubit quantum device now, the D-Wave 2000Q machine. %\cite{dwave}. 
This machine is not a circuit-based quantum computer. Rather it is what we call a quantum annealer; it solves optimization problems using a different method than execution of a quantum circuit, and it often solves these problems  successfully. 

But as of now we don't have a convincing theoretical argument or persuasive experimental evidence indicating that quantum annealers can really speed up the time to solution compared to the best classical hardware running the best algorithms for the same problems \cite{dwave,dwave2}. The situation is a bit nuanced. The quantum annealer is the noisy version (with rather poor quality qubits, in the case of the D-Wave machine) of what we call adiabatic quantum computing \cite{lidar}. We actually do have, for the case of noiseless qubits, a theoretical argument showing that adiabatic quantum computing is as powerful as circuit-based quantum computing \cite{aharonov}.  However, the argument demonstrating the equivalence of adiabatic and circuit-based quantum computation applies only if the adiabatic method is executed in a form that requires a high overhead cost in additional physical qubits, which is quite different from the way quantum annealers are used today. In any event, this formal argument applies only to noiseless qubits, and (in contrast to circuit-based quantum computing), we don't have a good theoretical argument attesting that quantum annealers are scalable. That is, for the case of noisy qubits we don't know how to show that quantum annealers will continue to work successfully as the problem size increases, even if we are willing to accept a sizable overhead cost to enforce robustness against noise. 

So far quantum annealers have been applied mostly to cases where the annealing is \textit{stoquastic} --- that means it might be relatively easy for a classical computer to simulate what the quantum annealer is doing \cite{stoquastic,jordan-stoquastic}. What's coming soon are non-stoquastic quantum annealers, which may have greater potential for achieving speedups over what the best classical algorithms can do. 

Since theorists have not settled whether quantum annealing is powerful, further experiments are needed. Experiments with quantum annealers over the next few years are likely to be quite informative. In particular, aside from the applications of quantum annealers to classical optimization problems, applications to quantum simulation problems should also be explored \cite{harris}. 

\subsection{Noise-resilient quantum circuits}

I've emphasized that eventually we are going to use quantum error correction to extend the size of the computations we can execute reliably using noisy quantum computers. But, because of the high overhead cost of quantum error correction, NISQ devices will not make use of it in the near term. Nevertheless, methods for mitigating the effects of noise may be an important consideration during the NISQ era. 

Naively one might expect that, for a generic circuit with $G$ gates, a single fault anywhere in the circuit could cause the computation to fail; if so, then we cannot execute the circuit reliably if the error rate per gate is much larger than $1/G$. However, at least for some problems and algorithms, that conclusion might be too pessimistic. 

In particular, for some of the quantum simulation algorithms of interest to physicists, there are only a relatively small number of circuit locations where a faulty gate can cause the computation to fail badly. This feature could arise because the circuit has low depth, and we can tolerate a constant probability of error per qubit in the final measurement outcome, or because the damage caused by an error occurring early in the circuit decays away by a later time. Such noise resilience is characteristic of some tensor network constructions which can be invoked for solving quantum optimization problems using Variational Quantum Eigensolvers \cite{kim,swingle}. 

There is a substantial opportunity for experimentalists and theorists, working together over the next few years, to find better ways of making quantum circuits noise resilient, and so extend the computational reach of NISQ technology. We should be wary, though, of a potential tradeoff --- making a quantum circuit more noise resilient may also make it easier to simulate classically. 

\subsection{Quantum deep learning}

Machine learning is transforming technology and having a big impact on science as well, so it is natural to wonder about the potential of combining machine learning with quantum technology. There are a variety of different notions of ``quantum machine learning.'' Much of the literature on the subject builds on quantum algorithms that speed up linear algebra and related tasks \cite{qml,aaronson-fine-print}, and I'll address such applications in the ensuing subsections. 

First, though, I'll comment on the potential of quantum deep learning \cite{duan}. Deep learning itself has become a broad subject \cite{hinton}, but to be concrete let us contemplate a restricted Boltzmann machine. This may be regarded as a spin system in thermal equilibrium at a low but nonzero temperature, with many hidden layers of spins separating the input and output. (The word ``restricted'' means that there are no couplings among spins within a given layer; instead only spins in successive layers are coupled.) The system may have millions of coupling parameters, which are optimized during a training phase to achieve a desired joint probability distribution for input and output. The training may be supervised, for example the network might be taught to recognize a labeled set of photos, or unsupervised, in which case the goal is to identify patterns in an unlabeled training set of data. The quantum analog of such a machine could have a similar structure and function, but where the spins are qubits rather than classical bits. It may be that quantum deep learning networks have advantages over classical ones; for example they might be easier to train for some purposes. But we don't really know --- it's another opportunity to try it and see how well it works. 

One possible reason for being hopeful about the potential of quantum machine learning rests on the concept known as QRAM --- quantum random access memory. By QRAM we mean representing a large amount of classical data very succinctly, by encoding a vector with $N$ components in just $\log N$ qubits. But typical proposals for quantum machine learning applications are confounded by severe input/output bottlenecks. For applications to large classical data sets one should take into account the cost of encoding the input into QRAM, which can nullify the potential advantages. Furthermore, the output from a quantum network is itself a quantum state; to get useful information we need to measure the state, and we can acquire no more than $\log N$ bits of classical information in a single-shot measurement of $\log N$ qubits. 

These bottlenecks arise when we try to train a quantum network to learn about correlations in classical data. Perhaps it is more natural to think about quantum machine learning in a setting where both the input and the output are quantum states. Hence we might expect a quantum network to have advantages over a classical one when used, for example, to control a complex quantum system. More broadly, it seems plausible that quantum deep learning would be more powerful than its classical counterpart when trying to learn probability distributions for which quantum entanglement has a significant role. That is, quantum deep learning networks might be very well suited for quantum tasks, but for applications of deep learning that are widely pursued today it is unclear why quantum networks would have an advantage. 

On the encouraging side, it's worth noting that a quantum deep learning machine won't necessarily need to be a general purpose circuit-based quantum computer. It might be a special purpose device --- perhaps a quantum annealer, if not too noisy.

\subsection{Quantum matrix inversion}

QRAM has further implications for quantum algorithms. In particular, the task of matrix inversion admits an exponential quantum speedup, which could have many applications. 

The algorithm we call HHL \cite{hhl} (after the three authors who first discussed it nine years ago) takes as input a succinct representation of a very large $N{\times}N$ matrix $A$, which has to be sufficiently sparse and well conditioned, and also as input an $N$-component vector $|b\rangle$,  encoded in QRAM as a quantum state of $\log N$ qubits. The output of the quantum algorithm is (with a small error) the quantum state $|A^{-1} b\rangle$, the result of applying the inverse of the matrix $A$ to the input vector $|b\rangle$. This quantum algorithm runs in time $O(\log N)$, an exponential speedup relative to classical matrix inversion (for fixed error and for fixed sparsity and condition number of $A$).

However, aside from the caveat that the matrix $A$ must be sparse and well conditioned, note that both the input vector $|b\rangle$ and the output vector $|A^{-1}b\rangle$  are quantum states. Once we have the output state, we can perform measurements to learn features of the output vector $A^{-1}b$, and by repeating the scheme many times we can learn about the output vector in greater detail. But if we are trying to apply this matrix inversion algorithm to classical input data, we need to take into account the potential cost of loading that classical data into QRAM, which could nullify the exponential speedup. Alternatively, loading classical data into QRAM might be avoided by computing $|b\rangle$ on a quantum computer rather than entering it from a database.
 
We do have good reason to believe this quantum matrix inversion algorithm is powerful, because it solves what we call a BQP-complete problem. That is, any problem that can be solved efficiently with a quantum computer can be encoded as an instance of this matrix inversion problem. And a number of potentially useful applications have been suggested. For example, we can use this matrix inversion algorithm to find approximate solutions to classical linear field equations \cite{clader,montanaro-pallister,jordan-wave} (after discretizing the field on a spatial lattice, and applying a suitable preconditioner). This procedure could be used, say, to solve the equations of electromagnetism in a complex three-dimensional geometry, for the purpose of optimizing the performance of an antenna.

I expect HHL to have high-impact applications eventually, but it is not likely to be feasible in the NISQ era.  The algorithm may just be too expensive to be executed successfully by a quantum computer which doesn't use error correction. 

\subsection{Quantum recommendation systems}

Another recently proposed quantum algorithm \cite{kerenidis} provides an exponential speedup over the best currently known classical algorithm for a useful task: making high-value \textit{recommendations}. The goal is to recommend a product to a customer that the customer will probably like, based on limited knowledge of the preferences of that customer and other customers. 

To formulate a simplified version of the problem, consider $m$ customers and $n$ products, and let $P$ be a binary $m\times n$ \textit{preference matrix} such that $P_{ai} = 1$ if customer $a$ likes product $i$ and $P_{ai}=0$ if customer $a$ dislikes product $i$. In practice we might have $m\approx 10^{8}$ users and $n\approx 10^6$ products, but the rank $k$ of the matrix is relatively small, perhaps $k\approx 100$. This small rank indicates that there are only a limited number of customer types; therefore, once we know a few of the preferences of a new customer,  we can accurately recommend other products that customer will like.

An algorithm for recommendation systems has two stages. In the first stage, performed offline, a low rank approximation to the preference matrix $P$ is constructed. In the second stage, performed online, a new customer reveals some preferences, and the system outputs a recommendation which that customer will like with high probability. It is this second online stage that admits a quantum speedup. The quantum runtime is $O( \textrm{poly}(k)\textrm{polylog}(mn) )$, while the best classical algorithm known requires time $\textrm{poly}(mn)$ to return a high-value recommendation. Crucially, the quantum algorithm, in contrast to the classical one, does not attempt to reconstruct the full recommendation matrix, which would take too long. Instead it samples efficiently from the low-rank approximation to $P$.

This is a significant quantum speedup for a real-world application of machine learning, encouraging the hope that other such speedups will be discovered. However, we don't currently have a convincing theoretical argument indicating that the task performed by quantum recommendation systems (returning a high-quality recommendation in polylog($mn)$ time) is impossible classically.\footnote{Note added: After this paper was submitted, Tang \cite{tang} exhibited a ``quantum-inspired'' classical algorithm which returns a high-value recommendation in runtime  $O( \textrm{poly}(k)\textrm{polylog}(mn) )$;  thus the quantum algorithm proposed in \cite{kerenidis} does not in fact achieve an exponential speedup compared to the best classical algorithm performing the same task.} Formulating such an argument would strengthen our confidence that quantum computers will have a significant impact on machine learning. Here too, though, the algorithm is probably too costly to be convincingly validated during the NISQ era. 

\subsection{Quantum semidefinite programming}
Semidefinite programming is the task of optimizing a linear function subject to matrix inequality constraints. Specifically, given as input the $m{+}1$ $N{\times}N$ Hermitian matrices $\{C, A_1, A_2, \dots, A_m\}$ and real numbers $\{b_1, b_2, \dots, b_m\}$, the problem is to find a positive semidefinite $N\times N$ matrix $X$ that maximizes $\textrm{tr}\left(CX\right)$ subject to the $m$ constraints $\textrm{tr}\left(A_iX\right)\le b_i$. Convex optimization problems of this type have widespread applications, and can be solved classically in $\textrm{poly}(m,N)$ time.

A recently discovered quantum algorithm \cite{brandao1,brandao2} finds an approximate solution to the problem in time $\textrm{polylog}(N)$, an exponential speedup. In the quantum algorithm, the output is a quantum state, a density operator $\rho$ which approximates the optimal matrix $X$. This quantum state can then be measured to extract information about $X$, and the procedure can be repeated many times to learn about $X$ is more detail.

However, in the analysis of the quantum algorithm one assumes the availability of an initial quantum state of a particular form, the thermal Gibbs state associated with a Hamiltonian which is a linear combination of the input matrices for the semidefinite program. Whether the exponential speedup can actually be realized, then, depends on whether this Gibbs state can be prepared efficiently using a quantum computer. 

For semidefinite programs with the right structure, the efficient preparation of the Gibbs state is possible, and therefore the exponential speedup really is feasible. In particular, the quantum algorithm achieves an exponential speedup if all the input matrices for the semidefinite program have low rank, or if the corresponding Hamiltonian system thermalizes rapidly. It is not yet clear to what extent these properties apply to semidefinite programs of practical interest; clarifying that issue is an important goal for further research.

It is intriguing that the crucial step in the quantum algorithm is the preparation of a thermal state at a nonzero temperature. That may suggest that the algorithm has some intrinsic robustness against thermal noise, or that the thermalization stage of the algorithm could be executed by quantum annealing. Conceivably, then, a quantum solver for semidefinite programs might be within reach of NISQ technology.

\subsection{Quantum simulation}

As I have already emphasized, we expect quantum computers to be very well suited for studying the properties of highly entangled systems of many particles. As Laughlin and Pines \cite{laughlin} emphasized in the statement I quoted earlier, we think that simulating such ``strongly correlated'' matter is a difficult computational problem, because many very good physicists have tried to solve it for decades and have not succeeded. Thus, as Feynman argued \cite{feynman}, the simulation of highly entangled quantum matter is the natural arena where quantum computers seem to have a clear advantage over classical ones.

In the long term, quantum simulation using quantum computers has great potential to wield a broad influence on the world \cite{microsoft,microsoft2,nsf-chem}. Advances in quantum chemistry, facilitated by quantum computing, might precipitate the design of new pharmaceuticals, and new catalysts that improve the efficiency of nitrogen fixation or carbon capture. New materials arising from quantum simulations might lead to more effective power transmission or improved collection of solar energy. We probably lack the imagination to anticipate the most transformative discoveries which will flow from more powerful quantum simulators. But this promise is not so likely to be fulfilled in the NISQ era, because quantum algorithms for accurately simulating large molecules and exotic materials may be too expensive to succeed without error correction. 

Classical computers are especially bad at simulating \textit{quantum dynamics} --- that is, predicting how a highly-entangled quantum state will change with time. Quantum computers have a big advantage for that task, and physicists hope to learn interesting things about quantum dynamics using NISQ technology in the relatively near term. It is instructive to recall that the theory of classical chaos (the extreme sensitivity to initial conditions in classical dynamical systems, which accounts for our inability to predict the weather more than two weeks out) advanced rapidly in the 60s and 70s after it became possible to simulate chaotic dynamical systems using classical computers. We may anticipate that the emerging ability to simulate chaotic quantum systems (those in which entanglement spreads very rapidly) will promote advances in our understanding of quantum chaos. Valuable insights might already be gleaned using noisy devices with of order 100 qubits. 

%%%
%[Comment on the quantum dynamics simulation task.]
%%%

\subsection{Digital vs. analog quantum simulation}

Having emphasized the potential value of simulating quantum dynamics, I should comment about the distinction between analog and digital quantum simulation. When we speak of an \textit{analog quantum simulator} we mean a system with many qubits whose dynamics resembles the dynamics of a model system we are trying to study and understand. In contrast, a \textit{digital quantum simulator} is a gate-based universal quantum computer which can be used to simulate any physical system of interest when suitably programmed, and can also be used for other purposes.

Analog quantum simulation has been a very vibrant area of research for the past 15 years, while digital quantum simulation with general purpose circuit-based quantum computers is just now getting started. But some of the same experimental platforms, for example trapped ions and superconducting circuits, can be used for both purposes. Analog quantum simulators have been getting notably more sophisticated, and are already being employed to study quantum dynamics in regimes which may be beyond the reach of classical simulators \cite{lukin,monroe}. But analog quantum simulators are hampered by imperfect control --- the actual quantum system in the lab only crudely approximates the model system of interest. For that reason, analog simulators are best suited for studying features that physicists call \textit{universal}, properties which are relatively robust with respect to introducing small sources of error. A major challenge for research using analog quantum simulators is identifying accessible properties of quantum systems which are robust with respect to error, yet are also hard to simulate classically.

We can anticipate that analog quantum simulators will eventually become obsolete. Because they are hard to control, they will be surpassed some day by digital quantum simulators, which can be firmly controlled using quantum error correction. But because of the hefty overhead cost of quantum error correction, the reign of the analog quantum simulator may persist for many years. Therefore, when seeking near-term applications of quantum technology, we should not overlook the potential power of analog quantum simulators. 

\subsection{Quantum games}

Advances in classical computing launched a new world of digital games, touching the lives of millions and generating billions in revenue. Could quantum computers do the same?

Physicists often say that the quantum world is counter-intuitive because it is so foreign to ordinary experience. That's true now, but might it be different in the future?  Perhaps kids who grow up playing quantum games will acquire a visceral understanding of quantum phenomena that our generation lacks. Furthermore, quantum games could open a niche for quantum machine learning methods, which might seize the opportunity to improve game play in situations where quantum entanglement has an essential role.

\section{The daunting climb to scalability}

I've already emphasized repeatedly that it will probably be a long time before we have fault-tolerant quantum computers solving hard problems. Nevertheless, we can make significant progress in the near term by developing better methods and hardware for implementing quantum error correction, guided by relatively small-scale experiments with quantum error-correcting codes. We expect that, in the next few years, enhanced control of an error-corrected qubit will be demonstrated convincingly in the lab for the first time. Further near-term experiments exploring quantum error correction are bound to advance our understanding of noise in quantum devices, and to suggest more effective techniques for noise mitigation \cite{terhal,emerson} and device validation \cite{flammia}.

Nevertheless, solving really hard problems (like factoring numbers which are thousands of bits long) using fault-tolerant quantum computing is not likely to happen for a while, because of the large number of physical qubits needed. To run algorithms involving thousands of protected qubits we'll need a number of physical qubits which is in the millions, or more \cite{fowler}.  That's a very large leap from where we will be for the next few years, with of order a hundred qubits. Crossing the ``quantum chasm,'' from hundreds of physical qubits to millions of physical qubits, is going to take some time, but we'll get there eventually.

Throughout the NISQ era it will remain important to strive for lower gate error rates in the various quantum platforms. With more accurate quantum gates, quantum computers operating without error correction will be able to go further by executing larger circuits. Better gates will also lower the overhead cost of doing quantum error correction when we eventually move on to fault-tolerant quantum computing in the future. (Microsoft's current program \cite{microsoft-topo}, focused on advancing topological quantum computing, is founded on the notion that much lower gate error rates will pay off handsomely sometime in the future.)

It is important to realize that we will need significant advances --- in basic science as well as in systems engineering --- to attain fully scalable fault-tolerant quantum computers. Therefore, the challenge of scalability poses compelling problems for scientists and engineers. Because we still have so far to go, new insights, developments, and innovations have the potential to substantially alter the long-term outlook.

\section{Summary}

This article is the result of an uneasy compromise. I hope the content provides a useful overview for technically informed readers, but I've tried to minimize jargon and technical details which could be distracting or off-putting for a broader audience. I have not attempted to compile a comprehensive list of references; instead I have referred to just a few papers that amplify or clarify points I wanted to make, or that might provide instructive further reading. I apologize to many colleagues whose highly valuable contributions are not properly acknowledged here. 

Let me now summarize  the main points I've tried to convey.

\begin{description}

\item %\noindent
Now is a privileged time in the history of science and technology, as we are witnessing the opening of the NISQ era (where NISQ = noisy intermediate-scale quantum). We'll soon have the opportunity to experiment with NISQ  technology to see what it can do. Perhaps NISQ will allow us to speed up the time to solution for problems of broad interest in the near future, but we don't know yet whether that will happen.

\item 
We have some specific ideas of things we'd like to try, such as hybrid quantum-classical algorithms for solving optimization problems, both classical and quantum ones. 

\item 
We can expect that, once we have quantum computers to experiment with,  the development of quantum algorithms will accelerate. Perhaps new heuristics will be discovered for solving useful problems, although we may not be able to explain why these heuristics work so well, at least not right away.

\item 
Although we probably won't be able to use full-blown quantum error correction to protect NISQ-era devices from noise, we should design near-term quantum algorithms with noise resilience in mind.  Clever design of noise-resilient algorithms may extend the computational power of NISQ devices. 

\item 
Classical computers are especially bad at simulating the dynamics of highly entangled many-particle quantum systems; therefore quantum dynamics is a particularly promising arena where quantum computers may have a significant advantage over classical ones.

\item 
We should continue to focus on building quantum hardware with lower gate error rates. Improved quantum gate accuracy will lower the overhead cost of quantum error correction when it is eventually implemented. In the nearer term, more accurate gates will allow us to execute larger quantum circuits, and so extend the power of NISQ technology.

\item 
NISQ is not likely to change the world all by itself. Instead, the primary goal for near-term quantum platforms should be to pave the way for bigger payoffs which will be realized by more advanced quantum devices down the road.

\item 
The truly transformative quantum technologies of the future are probably going to have to be fault tolerant, and because of the very hefty overhead cost of quantum error correction, the era of fault-tolerant quantum computing may still be rather distant. No one really knows how long it will take to get there. As the quantum community eagerly seizes the impending opportunity to experiment with NISQ devices, we must not lose sight of the essential longer-term goal: hastening the onset of the fault-tolerant era. 
\end{description}

I've adopted a cautionary stance in these remarks, because I think we should face the onset of the NISQ era with well-grounded expectations. But please don't misunderstand me --- like many of my colleagues, I'm eagerly anticipating the discoveries that will ensue as the NISQ era unfolds over the next few years. Quantum technology is rife with exhilarating opportunities, and surely many rousing surprises lie ahead. But the challenges we face are still formidable. All quantumists should appreciate that our field can fulfill its potential only through sustained, inspired effort over decades. If we pay that price, the ultimate rewards will more than vindicate our efforts. 

\section*{Acknowledgments}
This article is based on a Keynote Address delivered at Quantum Computing for Business on 5 December 2017.
I thank Matt Johnson %of QC Ware 
for organizing this stimulating meeting and inviting me to participate. My remarks here have been influenced by discussions with many colleagues, too many to list. But I've especially benefited from insights due to Scott Aaronson, Sergio Boixo, Fernando Brand\~ao, Elizabeth Crosson, Toby Cubitt, Eddie Farhi, Steve Flammia, David Gosset, Daniel Gottesman, Stephen Jordan, Jordan Kerenidis, Isaac Kim, Seth Lloyd, Shaun Maguire, Oskar Painter, David Poulin, Peter Shor, Brian Swingle, Matthias Troyer, Umesh Vazirani, and Thomas Vidick. Some of this work was done while I attended the 2017 program on Quantum Physics of Information at the Kavli Institute for Theoretical Physics (KITP).
%Of course, none of them are responsible for any of my false or misleading statements. 
I gratefully acknowledge support from ARO, DOE, IARPA, NSF, and the Simons Foundation. The Institute for Quantum Information and Matter (IQIM) is an NSF Physics Frontiers Center.
%with support from the Gordon and Betty Moore Foundation. 

\end{document}